\documentclass[%
reprint,
 amsmath,amssymb,
 aps,
 prl,
floatfix,
]{revtex4-2}
\usepackage{graphicx,xcolor}
\definecolor{darkblue}{RGB}{0,0,150}
\definecolor{nightblue}{RGB}{0,0,100}

\usepackage{mathrsfs,dsfont,mathtools}
\usepackage{dcolumn}% Align table columns on decimal point
\usepackage{bm}% bold math
\usepackage{pifont}
\usepackage[
colorlinks,
citecolor=darkblue,
linkcolor=darkblue,
urlcolor=nightblue]{hyperref}% add hypertext capabilities
\usepackage[english]{babel}
\usepackage[babel,kerning=true,spacing=true]{microtype}

\usepackage{feynmp-auto}

\definecolor{DarkRed}{RGB}{100,0,0}

\AtBeginDocument{\renewcommand{\natexlab}[1]{#1}}% <--- the fix
\newcommand{\cross}{\ding{55}}

\begin{document}

\title{Unification of Nonlinear Anomalous Hall Effect and Nonreciprocal Magnetoresistance in Metals
by the Quantum Geometry }

\author{Daniel Kaplan}
\author{Tobias Holder}
\author{Binghai Yan}
\email{binghai.yan@weizmann.ac.il}
\affiliation{Department of Condensed Matter Physics,
Weizmann Institute of Science,
Rehovot 7610001, Israel}

\date{\today}

\begin{abstract}
The quantum geometry has significant consequences in determining transport and optical properties in quantum materials. Here, we use a semiclassical formalism coupled with perturbative corrections unifying the nonlinear anomalous Hall effect (NLAHE) and nonreciprocal magnetoresistance (NMR, longitudinal resistance) from the quantum geometry. In the dc limit, both transverse and longitudinal nonlinear conductivities include 
a term due to the normalized quantum metric dipole. 
The quantum metric contribution is intrinsic and does not scale with the quasiparticle lifetime. 
We demonstrate the coexistence of a NLAHE and NMR driven by the quantum metric dipole in films of the doped antiferromagentic topological insulator
MnBi$_2$Te$_4$. 
Our work indicates that both longitudinal and transverse nonlinear transport provide a sensitive probe of the quantum geometry in solids.

\end{abstract}

\maketitle

\paragraph{Introduction.---}
The quantum geometry of wave functions significantly impacts  transport properties in quantum materials~\cite{shapere1989geometric,bohm2003geometric,Xiao2010}. It is encoded in the quantum geometric tensor~\cite{Provost1980} which includes the Berry curvature~\cite{berry1984quantal} and a quantum metric~\cite{Provost1980,berry1989quantum,Vanderbilt.Marzari1997}. As it is well known, the Berry curvature causes the intrinsic anomalous Hall effect in magnetic materials~\cite{Jungwirth.Niu.MacDonald2002,Onoda.Nagaosa2002}. In similar vein, it has been suggested that the Berry curvature dipole~\cite{Sodemann2015} and the quantum metric~\cite{Gao2014} generate a nonlinear anomalous Hall effect (NLAHE). The former was  predicted~\cite{Zhang2018a,Zhang2018b} and shortly after realized experimentally in Weyl semimetals (e.g. WTe$_2$ and MoTe$_2$)~\cite{Ma2019,Kang2019,tiwari2021giant}. More recently, the latter has been predicted to appear in antiferromagnetic metals (e.g., CuMnAs)~\cite{Yang.Liu2021,Xiao.Wang2021,holder2021mixed}. The NLAHE might be useful in optoelectronic applications such as Terahertz detection and radio frequency rectification~\cite{Fu.Zhang2021terahertz,Yang.Kumar2021room,Xie.Lu.Du2021review}. 

Another nonlinear transport phenomenon which has recently gained attention is the nonreciprocal magnetoresistance (NMR) (also called electric magnetochiral anisotropy)~\cite{Rikken2001}. Here, the longitudinal resistance exhibits a second order correction that can be reversed by the magnetism or a magnetic field, acting as a magnetic diode. It is extensively studied in noncentrosymmetric or chiral materials~\cite{Tokura.Nagaosa2018,Tokura.Yasuda2020QAHE,Cobden.Zhao2020WTe2,Iwasa.Ideue2021review,Shi.Li2022,Gao.Zhang2022}. The NMR is believed to originate either from the inelastic scattering by magnons~\cite{Tokura.Yasuda2016} and spin clusters~\cite{Nagaosa.Ishizuka2020} or alternatively from the second-order Drude conductivity caused by an asymmetric band structure~\cite{Iwasa.Ideue2017,Yan.Liu2021chirality}. 
We note that previous theories based on the Berry curvature or quantum metric~\cite{Gao2014,Yang.Liu2021,Xiao.Wang2021,zhuang2022extrinsic} lead to a vanishing second-order conductivity in the longitudinal direction. 

In this Letter, we propose a unification of the NLAHE and NMR in the same theory framework of second order perturbation theory, based on the quantum geometry. For the NLAHE, we find a significant interband correction related to the quantum metric. This quantum metric correction leads to a compact expression [i.e., Eq.~\eqref{eq:fullcurrent2}] for the nonlinear transport in metals and concomitantly predicts a NMR (i.~e. a non-vanishing longitudinal nonlinear conductivity). Here, this NMR is created partially by the normalized quantum metric dipole, besides by the known Drude term. We further provide a scaling relation with the linear conductivity ($\sigma_{||}$) to separate three contributions at second order; namely the 
(i) normalized quantum metric dipole (independent of $\sigma_{||}$) and
(ii) Berry curvature dipole (linear to $\sigma_{||}$) and  the (iii) Drude weight (quadratic to $\sigma_{||}$). These three contributions can be distinguished by symmetry restrictions, as summarized in Tab.~\ref{table:1}. 
We demonstrate the coexistence of a large NLAHE and the NMR in thin films of the well-studied, doped AFM topological insulator, MnBi$_2$Te$_4$ \cite{gong2019experimental}, and call for experimental verification. Our findings indicates that nonlinear longitudinal and transverse transports provide a sensitive probe for the quantum geometry in solids.

%We summarize our main findings by considering symmetry restrictions on the observation of the NLAHE and NMR, with particular application to thin films of MnBi$_2$Te$_4$ in Tab.~\ref{table:1}.

%\paragraph{Results and Discussions}

\begin{figure}
    \centering
    \includegraphics[width=\linewidth]{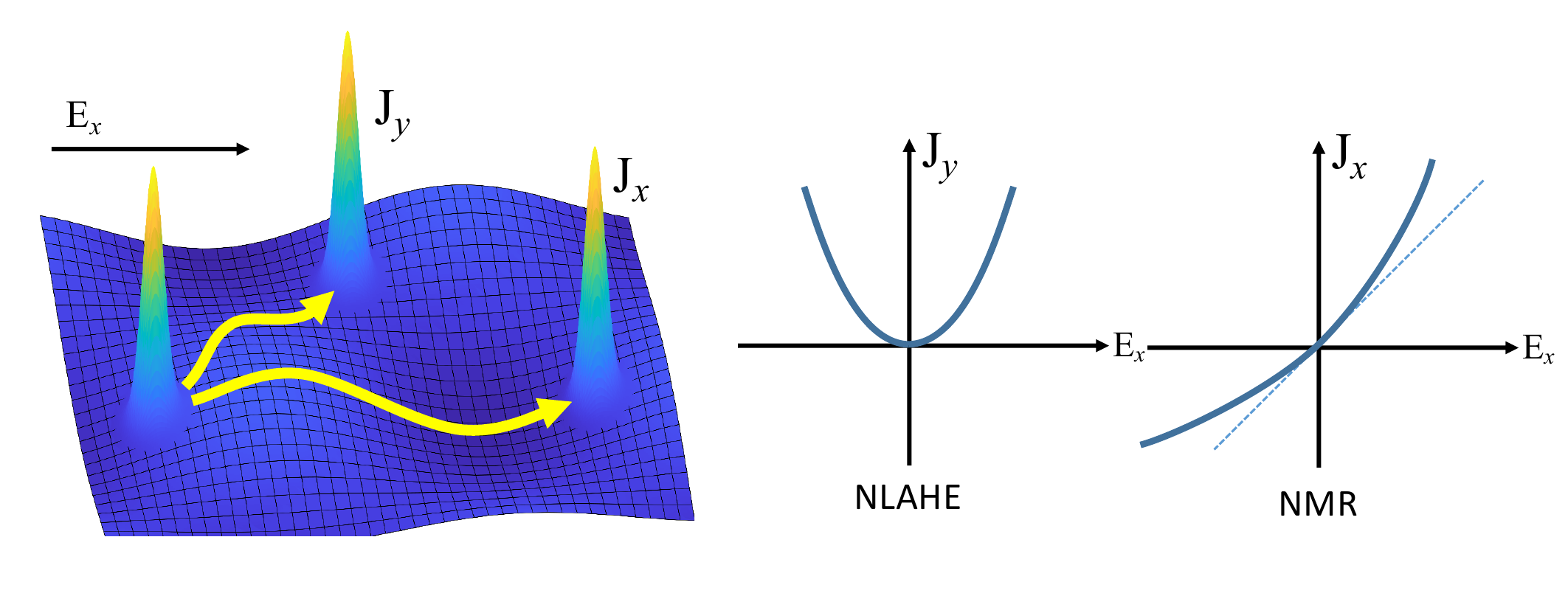}
    \caption{Schematics of the nonlinear anomalous Hall effect (NLAHE) and nonreciprocal magnetoresistance (NMR) due to the anomalous motion of wavepackets induced by the quantum geometry. The dependence of the current components $J_x$, $J_y$ with respect to the electric field $E_x$ are illustrated on the right.
    }
    \label{fig:1}
\end{figure}

\paragraph{Semiclassical kinetic equation.---}
We derive the nonlinear conductivity for a generic metal from a modification to the Boltzmann equation. Similar calculations have been performed very recently using the Kubo  formalism~\cite{Kaplan2022_scipost,Kaplan2022_ingap,Michishita2022}, where the manipulations can be done order by order in the lifetime. In the following, we show that by combining perturbation theory and the Boltzmann approach, such an analysis reveals that the renormalization of quasiparticle properties enables longitudinal nonlinear current response, stemming from the quantum geometry.

Within the semiclassical approach, the nonlinear conductivity arises from a renormalization of the distribution function and quasiparticle operators due to the applied field, consistent with other approaches. In the semiclassical approach, corrections which are indepedent of scattering time arise \textit{solely} due to correction to ground state quantities, while the scattering time enters through the time evolution of the distribution function.
For the electron distrbution function $f(\mathbf{r}, \mathbf{k}, t)$, the Boltzmann equation is,
\begin{align}
\partial_t f + \frac{\mathbf{F}}{\hbar} \nabla_{\mathbf{k}} f + \mathbf{v} \nabla_{\mathbf{r}} f = \mathcal{I}(f).
\label{eq:boltzmann}
\end{align}
Here, $\mathbf{F}$ is the semiclassical force in the presence of electric field, $\mathbf{F} = e\mathbf{E}$, $\mathbf{v}$ is the band diagonal velocity given by $v_n^c (\mathbf{k}) = \partial_{k_c } \varepsilon_n(\mathbf{k})$, with $\varepsilon_n (\mathbf{k})$ being the energy of Bloch state $n$ with momentum $\mathbf{k}$. For a uniform perturbation, we drop the spatial gradient acting on the distribution function. $\mathcal{I}$ refers to the collision integral, and throughout this work we adopt the relaxation time approximation and set $\mathcal{I}(f) = -\frac{f-f_0}{\tau}$, where $\tau$ is the scattering time and $f_0$ is the Fermi Dirac distribution. 
The solution to the equation follows by an order-by-order (in electric field) expansion of the density, $f = f_0 + f_1 + f_2 +\ldots$. The zeroth order evolves trivially in time, as expected. 
Working in the frequency domain, the two relevant deviations from equilibrium are,
\begin{align}
    i\omega f_1 + \frac{e\mathbf{E}}{\hbar}\partial_{\mathbf{k}}f_0  = -\frac{f_1}{\tau},  ~ i\omega f_2 + \frac{e\mathbf{E}}{\hbar}\partial_{\mathbf{k}}f_1  = -\frac{f_2}{\tau}.
    \label{eq:boltz1}
\end{align}
As the current is given by $\bm{j} = -e\int_{\bm{k}} \bm{v} f $, it is imperative to consider correction to second order in the applied field not only in $f$ but also in $\bm{v}$.

We study these corrections using a modified form of the Luttinger-Kohn method \cite{Luttinger1955}. 
In the presence of an electric field, the band Hamiltonian $H_0$ is modified by the coupling term $H_1 = -e\mathbf{E}\cdot\mathbf{r}$, where $\mathbf{r}$ is the position operator. 
The effective low energy degrees of freedom are therefore given by a renormalized band Hamiltonian $H'$ which can be obtained through the unitary transformation $H' = e^{S}H e^{-S}$. To fix $S$, we expand $e^{S}H e^{-S}$ to first order in $\mathbf{E}$, with the condition,
\begin{align}
    H_1 + [S,H_0] = 0.
\end{align}
This yields the matrix elements
\begin{align}
    S_{nm} &= -e \frac{E_a\mathcal{A}^a_{nm}}{\varepsilon_{nm}}, ~ n\neq m,
\end{align}
where $S_{nn}=0$ and $\varepsilon_{nm} = \varepsilon_{n} - \varepsilon_{m}$. The inter-band Berry connection is defined as $\mathcal{A}^a_{nm} = \left\langle n\mathbf{k} | \hat{r} |m \mathbf{k}\right\rangle$. The immediate effect of the transformation $S$ is in the renormalization of operators. The diagonal part of the Berry connection is analogously transformed, $\mathcal{A} \to e^{S} \mathcal{A}^{-S}$,
\begin{align}
    \mathcal{A}'^{a}_{n} = \mathcal{A}^{a}_{n} + \left[S, \mathcal{A}^a\right]_{n} = \mathcal{A}^{a}_{n} - e E^b G^{ba}_{n}.
    \label{eq:berry_shift}
\end{align}
The energy is accordingly renormalized,
\begin{align}
    \varepsilon_n'= \varepsilon_n(\mathbf{k}) - e \mathbf{E}\cdot \mathbf{r}'_n = \varepsilon_n(\mathbf{k}) + e^2G^{ab}_{n} E_a E_b,
    \label{eq:energy_renorm}
\end{align}
where $G^{ab}_{n} = \sum_{m\neq n} \frac{\left(\mathcal{A}^a_{nm}\mathcal{A}^b_{mn} + \mathcal{A}^b_{nm}\mathcal{A}^a_{mn}\right)}{\varepsilon_{nm}}$ is the band-normalized
quantum metric and $\mathbf{r}'_n$ is the linear in $\mathbf{E}$ correction to $\mathcal{A}_n$, Eq.~\eqref{eq:berry_shift} (The full derivation of the correction to the Hamiltonian is found in the SI).
The velocity operator incurs both a first and second order correction via dressing with $S$, giving,
\begin{align}
    v'^{a}_{n} = \frac{\partial \varepsilon'_n}{\partial {k_a}} + [S, v^a]_{nn} = v'^a_{n} - e \mathbf{E} \times \Omega'_n. 
    \label{eq:anomvel}
\end{align}
The term linear in $E$ is the familiar anomalous velocity, which is itself corrected by the electric field.
The corrected Berry curvature is then obtained via,
\begin{align}
    \Omega'^{\alpha \beta} = \partial_{\alpha} \mathcal{A}'^{\beta}_{n} - \partial_{\beta} \mathcal{A}'^{\alpha}_{n}.
\end{align}

\paragraph{Scattering time analysis.---}
The solutions to Eq.~\eqref{eq:boltz1} give terms which depend explicitly on the scattering time $\tau$. Working iteratively, it is,
\begin{align} 
    f_1 = \frac{-e \mathbf{E} \nabla_{\mathbf{k}} f_0}{\hbar(i\omega + 1 / \tau)}, ~~    f_2 = \frac{e^2 E_a E_b}{\hbar^2(i\omega + 1 / \tau)^2}\frac{\partial f_0}{\partial k_a \partial k_b}.
    \label{eq:solboltz} 
\end{align}
We arrive at two charge current pieces along the $c$-direction at second order in the electric field which depend explicitly on the scattering time, 
\begin{align} 
\notag
j^c_{1} &= -\frac{e E_a E_b}{\hbar}\sum_{n}\int_{\mathbf{k}} \frac{e^2}{(i\omega + 1/ \tau)^2}\frac{\partial^2 f_{n}}{\partial k_a \partial k_b} \frac{\partial \varepsilon_n}{\partial k_c}, \\ 
j^c_{2} &=  \frac{e E_a E_b}{\hbar}\sum_{n}\int_{\mathbf{k}} \frac{e^2}{(i\omega + 1/ \tau)}\frac{\partial f_{n}}{\partial k_a}\Omega^{bc}_{n} + (a \leftrightarrow b).
\label{eq:semis}
\end{align}
Here, $j^c_{1}$ results from combining the second order correction to the density $f_2$ (right hand term in Eq.~\eqref{eq:solboltz}) with the unperturbed band velocity $v^c_n = \frac{\partial \varepsilon_n}{\partial k_c}$. $j^c_{2}$, similarly draws from the perturbed velocity in Eq.~\eqref{eq:anomvel} with the first order correction in $f_1$ (left hand term in Eq.~\eqref{eq:solboltz}).
Note that the sum is over all ground state single-particle bands $n$. Besides the terms generated by the perturbation of the semiclassical density ($j^c_{1,2}$), there are extra corrections to the operators. The associated currents are due to corrections to the dispersion ($j^c_{\textrm{disp}}$) and to the anomalous velocity ($j^c_{\textrm{anom}}$), respectively. We stress that although these currents couple to the equilibrium density, they exist only for finite electric field. They are given by,
\begin{align}
\notag
    j^c_{\textrm{disp}} &= -\frac{e E_a E_b}{\hbar}\int_{\mathbf{k}} f_0 v'^c_n = -\frac{e^3 E_a E_b}{\hbar}\sum_n \int_{\mathbf{k}}f_n \partial_{k_c} G^{ab}_n,  \\
    j^c_{\textrm{anom}} &= \frac{e^2}{\hbar}\int_{\mathbf{k}} \frac{f_0}{2}\left( E_a\Delta\Omega^{ac}_n+E_b\Delta\Omega^{bc}_n\right),
\end{align}
where in the last line the expressions are written manifestly symmetric with respect to $(a \leftrightarrow b)$. We also include the linear-in-field contribution to the Berry curvature with the notation $\Delta\Omega^{ac} = -\left(eE_a\partial_{k_a}G^{ba}_n- eE_b\partial_{k_c} G^{cb}_n\right)$. Assembling all terms ($j^c_{1},j^c_{2}, j^c_{\textrm{disp}}, j^c_{\textrm{anom}}$), on arrives at the conductivity, which is given by $j^{c} / E^a E^b$,
\begin{subequations}\label{eq:fullcurrent2}
\begin{align}
    %\notag & \frac{j^c}{E^a E^b} \equiv  
    \notag &\sigma^{ab;c} =
    \\ \label{eq:fullcurrent2-drude}
    & -\frac{e^3\tau^2}{\hbar^3}\sum_n \int_{\bm{k}} f_n \partial_{k^a} \partial_{k^b} \partial_{k^c} \varepsilon_{n} ~  \\ \label{eq:fullcurrent2-berry}
    & -\frac{e^3 \tau}{\hbar^2}\sum_n \int_{\bm{k}} f_n ( \partial_{k^a} \Omega^{bc}_n + \partial_{k^b} \Omega^{ac}_n)
    \\ \label{eq:fullcurrent2-metric}
    & -\frac{e^3}{\hbar} \sum_n  \int_{\bm{k}} f_n ( 2\partial_{k^c}G_n^{ab} - \tfrac{1}{2}\left(\partial_{k^a} G_n^{bc} +\partial_{k^b}G_n^{ac})\right).
\end{align}
\end{subequations}
We observe that $\partial_{k^a} G_n^{bc}$ is the band-normalized quantum metric dipole. 
Eq.~\eqref{eq:fullcurrent2-drude} refers directly to the nonlinear Drude weight \cite{Kaplan2022_scipost} and is obtained by integrating twice by parts the first line in Eq.~\eqref{eq:semis}.
Similarly, the Berry curvature dipole Eq.~\eqref{eq:fullcurrent2-berry} is obtained by integrating by parts the second line of Eq.~\eqref{eq:semis}, in agreement  with Ref.~\cite{Sodemann2015}. 
The intrinsic contribution, which is $\tau$-independent in Eq.~\eqref{eq:fullcurrent2-metric}, is caused by the band-normalized quantum metric dipole. The Fermi surface contribution to the current of this intrinsic term is obtained again by integrating by parts. 
It is should be noted that Eq.~\eqref{eq:fullcurrent2-metric} violates the ($a,b,c$) cyclic permutation symmetry, the source of which is a gravitational anomaly~\cite{holder2021mixed}. 

\paragraph{Discussion.---}
As demonstrated,  Eq.~\eqref{eq:fullcurrent2-metric} naturally decomposes into $j^c_{\textrm{disp}}$ which propagates with the current direction $\partial_{k^c} G_n^{ab}$, and the purely transverse current  $j^c_{\textrm{anom}}$, related to $\partial_{k^c} G_n^{ab} - \frac{1}{2}(\partial_{k^a} G_n^{bc} + \partial_{k^b} G_n^{ac})$ which vanishes whenever $a = b = c$~\cite{Gao2014,Gao2019}. 
Therefore, Eq.~\eqref{eq:fullcurrent2} naturally unifies both longitudinal and transverse effects at nonlinear order. 
Specifically, we find that the longitudinal component (i.e. the NMR) has contributions from both 
the Drude conductivity ($\sim\tau^2$) and the band-normalized quantum metric dipole, independent of $\tau$.  
On the other hand, the NLAHE is constituted by the Berry curvature dipole at order $\tau$, and again the band-normalized quantum metric dipole. 
The band-normalized quantum metric $G_{n}^{ab}$ thus enters in both the longitudinal and the transverse components. 

The second order conductivity has received a lot of attention lately, with some variation in the precise form of the expressions due to the large number of terms involved~\cite{Gao2014,Watanbe2020,Michishita2022,Oiwa2022}. 
While there is some disagreement between different approaches regarding coefficients, in all works there exists a broad consensus that the nonlinear conductivity is nonzero at all orders of $\tau$. 
A simple criterion to validate our result is that Eq.~\eqref{eq:fullcurrent2} respects the intrinsic permutation symmetry when exchanging $E_a$ and $E_b$, as required by the dc response. 
The common geometric origin of longitudinal and transverse components shows that unlike for the linear conductivity, at second order, valuable information about the band structure geometry is accessible in either spatial component.

We point out two possible limitations of our calculation.
Firstly, second order perturbation theory captures the instantaneous response of the system, but is insensitive to non-perturbative effects like a steady-state equilibration at long times.
Secondly, we focused on effects of the band structure in a system with finite quantum lifetime $\tau$. Further extrinsic contributions to the conductivity beyond the relaxation time approximation were neglected. In principle, the nonlinear Hall effect is composed of extrinsic contributions \cite{Du2019,Du2021} which induce corrections to the nonlinear Hall conductivity to all three contributions, at order $\tau^2$, $\tau$ and $\tau^0$. 
However, for a collinear antiferromagnet (such as MnBi$_2$T$_4$ discussed in the following section) skew scattering contributions are expected to be significantly suppressed \cite{Watanbe2020}. More exotic mechanisms such as anomalous skew scattering proposed by Ref.~\cite{Ma2022_anomalousskew} require breaking of $\mathcal{C}_{3z}$-symmetry to appear, and are therefore not relevant in MnBi$_2$Te$_4$.
\begin{figure*}[tbp]
    \centering
    \includegraphics[width=\linewidth]{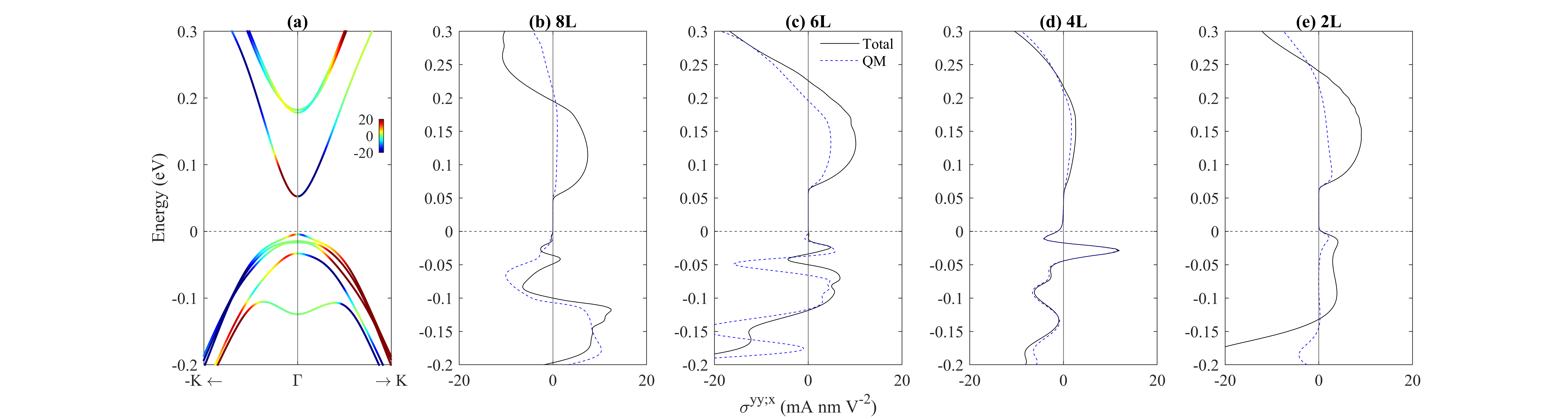}
    \caption{Band structure and nonlinear conductivity of antiferromagnetic MnBi$_2$Te$_4$ thin films ($\sigma^{yy;x} \equiv -\sigma^{xx;x}$ in this case).
    (a) The band structure of an eight-layer (8L) thick film. The band-decomposed contribution to $\sigma^{yy;x}$ is indicated by the color. (b)-(e) The energy dependence of $\sigma^{yy;x}$ for different films with the contribution from quantum metric (QM).
    }
    \label{fig:2}
\end{figure*}

To detect the signatures of the quantum geometry in experiments, we suggest examining the scaling relation between the transverse second order ($\sigma^{(2)}_{\perp}$) 
and longitudinal linear conductivity ($\sigma_{||}$), which reads
\begin{align}
     \sigma^{(2)}_{\perp} =
    \eta_2 (\sigma_{||}) ^2+ \eta_1 \sigma_{||} + \eta_{0}.
    \label{eq:scaling}
\end{align}
Here, the coefficients $\eta_i$ denote the respective part of the nonlinear conductivity which contributes at order $\mathcal{O}(\tau^{i})$. Namely, $\eta_2$ contains the nonlinear Drude term, and if present contributions from skew scattering~\cite{Du2021}. $\eta_1$ contains the Berry curvature dipole term, as well as extrinsic contributions related to side jump scattering. $\eta^{0}$ contains the effect of the normalized quantum metric. 
Similarly, for the longitudinal components it holds that

\begin{align}
   \sigma^{(2)}_{||} =
    \eta'_2 (\sigma_{||})^2 + \eta'_1 \sigma_{||} + \eta'_{0},
    \label{eq:scaling2}
\end{align}
where $\eta_1'$ will be nonzero only if there are extrinsic contributions.
Using these scaling relations, Eqs.~(\ref{eq:scaling},\ref{eq:scaling2}), it is therefore possible to isolate each of these terms and to quantify both the NLAHE and the NMR.

\paragraph{Realization in a Magnetic Metal.---}

In Eq.~\eqref{eq:fullcurrent2}, the Drude term and the normalized quantum metric dipole are antisymmetric under momentum inversion ($k\rightarrow -k$) which is related to the inversion symmetry ($\mathcal{P}$) or time-reversal symmetry ($\mathcal{T}$). 
Thus, breaking both $\mathcal{P}$ and $\mathcal{T}$ is required to obtain a nonzero contribution over the full Brillouin zone. In contrast, the Berry curvature dipole integral requires breaking both $\mathcal{P}$ and the combined symmetry $\mathcal{PT}$ (see Table \ref{table:1}).

While the NLAHE induced by the Berry curvature dipole was already observed in many materials, it is much harder to realize the quantum metric dipole-driven NLAHE or likewise the NMR. We will now demonstrate the coexistence of both NLAHE and NMR in a doped antiferromagnetic (AFM) topological insulator, MnBi$_2$Te$_4$. MnBi$_2$Te$_4$ is a layered van der Waals material with the A-type AFM structure. Thin film of this material with an even number of layers break $\mathcal{P}$ and $\mathcal{T}$, but preserve $\mathcal{PT}$. The $\mathcal{PT}$-symmetry specifically excludes the Berry curvature dipole contribution in the NLAHE, so that we can focus on the effect of the quantum metric. 
Experimentally, $\mathcal{PT}$ seems to be weakly broken in some MnBi$_2$Te$_4$ samples. For example, $\mathcal{PT}$-breaking was witnessed by the finite anomalous Hall signal for a six-layer-thick film  \cite{Xu.Ovchinnikov2021MBT}. 
However, even in this case the Berry dipole contribution to the NLAHE is still strongly suppressed by the three-fold rotational symmetry ($\mathcal{C}_{3z}$).

The crystal symmetry helps us understand the shape of nonlinear conductivity tensor. 
In a 2D film of MnBi$_2$Te$_4$, we set $x$ along the lattice vector direction in the basal plane for convenience. Here, $\mathcal{C}_{3z}$ constrains that the NLAHE and NMR share the same amplitude but opposite sign, i.e. it holds that
\begin{align} \notag
    \sigma^{yy;x} = -\sigma^{xx;x}, \\ \notag
    \sigma^{xx;y} = -\sigma^{yy;y}.
\end{align}
The combined symmetry~\cite{Yan.Tan2022MBT} by mirror reflection ($\mathcal{M}_x, x\rightarrow-x)$ and $\mathcal{T}$ then enforces $\sigma^{xx;y} = -\sigma^{yy;y} = 0$. Therefore, we only have one independent nonlinear conductivity, the NLAHE conductivity $\sigma^{yy;x}$, or equivalently the NMR conductivity $\sigma^{xx;x}$. When rotating the sample, we obtain the angle ($\theta$) dependence in the new coordinates ($x^\prime,y^\prime$) by $\sigma^{y^\prime y^\prime; x^\prime} = -\sigma^{x^\prime x^\prime; x^\prime} =\cos(3\theta) \sigma^{yy;x}$

Figure \ref{fig:2} shows the band structure and $\sigma^{yy;x}$ calculated on a eight-layer (8L) thick film by first-principles methods. We carried out ab-initio density-functional calculations \cite{VASP} on slab models of 2,4,6,8 layered MnBi$_2$Te$_4$ with AFM order. We then projected the converged wavefunctions of each slab onto local Wannier functions \cite{wannier90} of Bi-$p$ and Te-$p$ orbitals, which accurately span the energy window around the Fermi level. The energy dispersion is asymmetric between $\Gamma-\Bar{K}$ and  $\Gamma-K$ because both $\mathcal{P}$ and $\mathcal{T}$ are broken. Each energy state is furthermore doubly degenerate due to $\mathcal{PT}$ symmetry. The lowest conduction bands and highest valence bands contribute opposite signs in the nonlinear conductivity. Between $\mathbf{k}$ and $-\mathbf{k}$, bands contribute opposite signs to $\sigma^{yy;x}$, but at different amplitude. Indeed, the asymmetry between $\mathbf{k}$ and $-\mathbf{k}$ leads to nonzero $\sigma^{yy;x}$. As varying the Fermi energy, $\sigma^{yy;x}$ shows sign changes when a group of new bands appear at the Fermi surface. It vanishes in the energy gap.
Upon increasing the number of layers from two to eight layers, the region near to the lowest conduction band exhibits comparatively small changes in $\sigma^{yy;x}$, while the valence band region changes dramatically (see Fig.~\ref{fig:2}c). This is related to the fact that the lowest conduction bands are composed of gapped Dirac surface states while the top valence states have a bulk origin~\cite{Yan.Tan2022facet}. We therefore conclude that films thicker than two layers have similar surface states in the lowest conduction bands. 

Numerically, $\sigma^{yy;x}$ is in the order of magnitude of several $\it m \rm A\cdot \it n \rm m\cdot V^{-2}$, when using a relaxation time $\tau = 0.04~\rm ps$ to evaluate the Drude weight. 
As MnBi$_2$Te$_4$ samples commonly suffer from defects and exhibit a low mobility~\cite{Wu.Huang2020}, the Drude contribution may be smaller in reality. The presence of defects may also change the magnitude of the conductivity due to the decreased band gap and renormalization of surface state dispersion. A calculation for a defective slab is presented in the SI. Given that similar transport devices have recently become readily available~\cite{Xu.Gao2021layer,Gao.Zhang2022},  MnBi$_2$Te$_4$ films are ideal candidates to explore the quantum metric-driven NLAHE and NMR.

\paragraph{Summary.---}
We have shown that signatures of the quantum geometry of the band structure are imprinted in the second order conductivity not only in the transverse components but also in the longitudinal ones. 
To this end, we derived a NLAHE and NMR which both appear due to the quantum metric, and explored their effect in thin films of the $\mathcal{PT}$-symmetric antiferromagnet MnBi$_2$Te$_4$.
We found an intrinsic contribution to both NLAHE and NMR in antiferromagnetic thick films, providing an ideal platform to detect the quantum metric.

\begin{table}[h!]
\caption{ Symmetry restrictions for three contributions in Eq.~\eqref{eq:fullcurrent2} for a two-dimensional system regarding the inversion symmetry ($\mathcal{P}$), time-reversal symmetry ($\mathcal{T}$) and combined $\mathcal{P}\mathcal{T}$ symmetry. 
With the rotational symmetries $C_{3z}$ related to MnBi$_2$Te$_4$ film, both longitudinal ($\mathbf{j}_{||}$) and transverse currents ($\mathbf{j}_{\perp}$) exists. We note that $\mathbf{j}_{||}$ is dissipative because $\mathbf{j}\cdot \mathbf{E} \neq 0$, despite that the quantum metric-induced $\mathbf{j}_{||}$ is $\tau$-independent.
}
\label{table:1}
% \centering
\begin{tabular}{c |c c c c c c} 
 \hline
 Mechanism & $C_{3z}$ & $\mathcal{P}$ & $\mathcal{T}$ & $\mathcal{P}\mathcal{T}$ & $\mathbf{j}_{||}$ & $\mathbf{j}_{\perp}$   \\ 
 \hline 
 Nonlinear Drude & \checkmark  & \cross
 & \cross & \checkmark & \checkmark & $ \checkmark $ \\ 
 \hline 
 Berry curvature dipole & \cross  & \cross
 & \checkmark & \cross & \cross & \checkmark \\  \hline
 Quantum metric & \checkmark & \cross & \cross
 & \checkmark & \checkmark & \checkmark \\
 \hline
\end{tabular}
\end{table}

Our results further strengthen the observation that nonlinear responses carry more intricate and and the same time much more interesting information about the quantum geometry than linear response functions~\cite{Holder2021cat,Ahn2021}. It is imperative to further explore these aspects systematically in theory, for example for finite frequency response functions, and also for the magneto-transport. At the same time, present sample quality and device technology have the capabilities to detect these phenomena experimentally.

During the review of our manuscript, the quantum metric-induced NLAHE and NMR were observed in MnBi$_2$Te$_4$ thin films by a very recent experiment~\cite{wang2023quantum}.

%\clearpage
\textit{Acknowledgements--}
We acknowledge helpful discussions with Prof. Weibo Gao from Nanyang Technological University and Prof. Yang Gao from University of Science and Technology of China. 
B.Y.\ acknowledges the financial support by the European Research Council (ERC Consolidator Grant No. 815869, ``NonlinearTopo'') and Israel Science Foundation (ISF No. 2932/21). 
D.K.\ acknowledges support from the Weizmann Institute Sustainability and Energy Research Initiative. 
\bibliography{main.bbl}
%\end{acknowledgements}
\onecolumngrid
\newpage
 \appendix
% %section{A}
% \label{app:A}

\setcounter{equation}{0}
\setcounter{figure}{0}
\setcounter{table}{0}
\setcounter{page}{1}
\makeatletter
\renewcommand{\theequation}{S\arabic{equation}}
\renewcommand{\thefigure}{S\arabic{figure}}
\renewcommand{\bibnumfmt}[1]{[S#1]}
\begin{center}
\textbf{Supplemental Material: Unification of Nonlinear Anomalous Hall Effect and Nonreciprocal Magnetoresistance in Metals
by the Quantum Geometry }
\end{center}
\section{Time dependent Schrieffer-Wolff transformation}
In this section we derive Eq. 2. We follow the formal aspects of the time-dependent Schrieffer-Wolff transformation presented in Ref.~\cite{Onishi2022}.
We start with a time dependent drive $\mathcal{E}(t) = \mathbf{E}e^{i\omega t}+\mathbf{E}^{*}e^{-i\omega t}$. The perturbation is given by $H_1 = -e \mathcal{E}\mathbf{r}$. The total Hamiltonian is $H = H_0 + H_1$. We then construct the time dependent transformation $H \to e^{S} H e^{-S} + e^{S}\partial_t e^{-S}$ and expand the Hamiltonian order by order in $S(t)$. In what follows, we focus on the rectified component ($\omega + (-\omega)$) and suppress the explicit time dependence. We keep terms to order $S^2$.
\begin{align}
H \approx H_0 - e\mathcal{E} \mathbf{r} + [S,H_0] + \partial_t S + [S, - e\mathcal{E} \mathbf{r}] + \frac{1}{2} \left\lbrace S^2, H_0\right\rbrace - S H_0 S + \frac{1}{2} [S, \partial_t S].
\label{eq:schriff_wolff2nd}
\end{align}
We project on two general bands $n,m$. The Kohn-Luttinger condition introduced in the main text becomes,
\begin{align}
    \left(\varepsilon_{mn} + \partial_t S_{nm}\right) = e\mathcal{E} \mathcal{A}_{nm}.
\end{align}
We Fourier transform the above equation, and arrive at the time-domain solution,
\begin{align}
    S_{nm} = -e \mathcal{A}_{nm} \cdot \left(\frac{\mathbf{E}e^{i\omega t}}{\omega - \varepsilon_{nm}}+\frac{\mathbf{E}e^{-i\omega t}}{-\omega - \varepsilon_{nm}}\right).
\end{align}
Inserting this into the remaining terms Eq.~\eqref{eq:schriff_wolff2nd}, and isolating the rectified component, 
\begin{align}
    H_{nm} \approx H_{nm, 0} + \frac{e^2}{2} E_a E_b \sum_{l \neq m,n}\left(\mathcal{A}^a_{nl}\mathcal{A}^b_{lm} \left(\frac{1}{-\omega + \varepsilon_{nl}}+\frac{1}{-\omega + \varepsilon_{ml}}\right)+\mathcal{A}^b_{nl}\mathcal{A}^a_{lm} \left(\frac{1}{\omega + \varepsilon_{nl}}+\frac{1}{\omega + \varepsilon_{ml}}\right)\right).
\end{align}
In the final step, we project onto band $n$, by calculating $H_{nn}$ and taking the $\omega \to 0$ limit. We find,
\begin{align}
\notag 
H_{nn} &\approx \varepsilon_n + \frac{e^2}{2} E_a E_b \sum_{l \neq n}\left(\mathcal{A}^a_{nl}\mathcal{A}^b_{ln} \left(\frac{1}{\varepsilon_{nl}}+\frac{1}{ \varepsilon_{nl}}\right)+\mathcal{A}^b_{nl}\mathcal{A}^a_{lm} \left(\frac{1}{\varepsilon_{nl}}+\frac{1}{\varepsilon_{nl}}\right)\right) = \\ &
 \varepsilon_n + e^2 E_a E_b \sum_{m\neq n}\left(
\frac{\mathcal{A}^a_{nm}\mathcal{A}^b_{mn}+\mathcal{A}^b_{nm}\mathcal{A}^b_{mn}}{\varepsilon_{nm}} \right) = \varepsilon_n + e^2 E_a E_b G^{ab}_n,
\end{align}
Giving us precisely Eq. 6 in the main text.
\section{Calculation on a defective 4L slab}

As thin films of MnBi\textsubscript{2}Te\textsubscript{4} (MBT) are known to exhibit point defects, and specifically $Mn-Bi$ site mixing \cite{Garnica2022,Yan.Tan2022facet,Wu2023,Tan2023.distinct}.
Surface defects can significantly suppress the surface magnetic gap, which may enhance the nonlinear conductivity.

We tested out a defective slab containing two nearest-neighbor surface Mn\textsubscript{Bi} anti-site defects, in a manner preserving $\mathcal{PT}$ symmetry. We first constructed a 2$\times$2 slab out of pristine 4SL MBT, and placed Mn\textsubscript{Bi} anti-site defects at the two and bottom SL. The structure was relaxed without SOC with the forces totaling less than $0.01 \textrm{eV}\AA^{-1}$. The band structure of the defective slab is presented in Fig.~S1b while the nonlinear conductivity is plotted in Fig.S1c with comparison to the magnitude of the pristine case. The calculation was carried out along the Fermi surface, by modifying Eq. 12(c) in the main text,
\begin{align}
    \sigma^{yy;x} = \frac{2\pi e^3}{\hbar}\int \textrm{d} \ell_{k} \frac{1}{|\nabla \varepsilon_n(\mathbf{k}_F)|} v^{x}_n(\mathbf{k}_F) G^{yy}_{n}(\mathbf{k}_F).
\end{align}
Here $\textrm{d} \ell_{k}$ is a line element along the Fermi surface.
\begin{figure}
    \centering
    \includegraphics[width=0.8\textwidth]{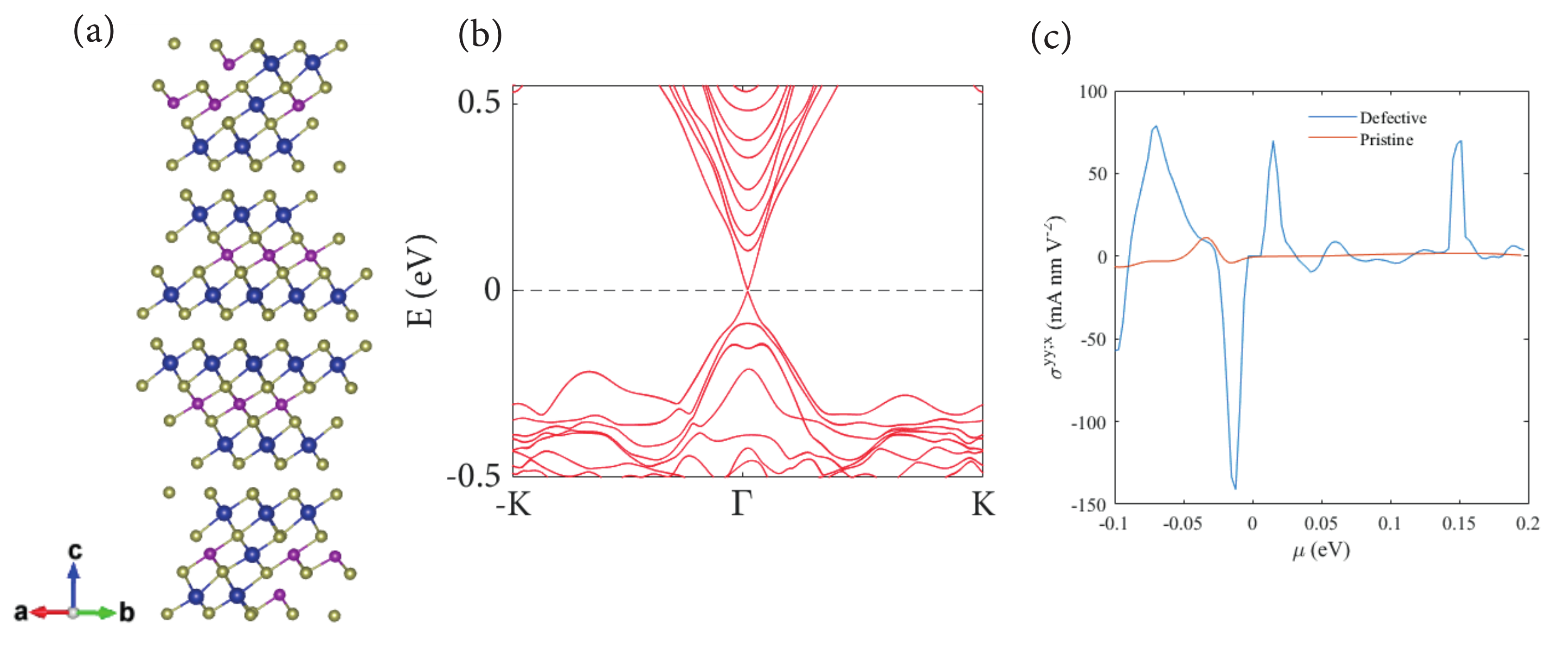}
    \caption{(a) Crystal structure of $\mathcal{PT}$ symmetric 2$\times$2 4SL MBT with 2 Mn\textsubscript{Bi} anti-site defects. Mn atoms are purple, while Bi is denoted with blue. (b) Band structure of the defective slab with a band gap of $E_g \approx 8 \textrm{meV}$. (c) Calculated $\sigma^{yy;x}$ (blue line) as compared with the pristine value (orange).}
    \label{figS1}
\end{figure}
\end{document}